\RequirePackage{fixltx2e}

\documentclass[twocolumn,showpacs,preprintnumbers,amssymb]{revtex4}
 
\usepackage{graphics}
\usepackage{epsfig}  
\usepackage{hyperref}
\usepackage{graphics}
\usepackage{color}      % use if color is used in text
\usepackage{graphicx}
\usepackage{graphicx}% Include figure files
\usepackage{dcolumn}% Align table columns on decimal point
\begin{document}

% Task  read about multiplicative noise, time dependent viscous medium,  space dependent viscous medium  write paragraphs. Introduce rescaled parameters for plot
% read more and every write for an hour,  correct grammar, typo etc....

\title { Effect  of viscous friction  on entropy,   entropy production and entropy extraction rates in underdamped and overdamped media}
\author{Mesfin Asfaw  Taye}
\affiliation {West Los Angles College, Science Division \\9000  Overland Ave, Culver City, CA 90230, USA}%Lines break automatically or can be forced with \\

\email{tayem@wlac.edu}

\begin{abstract}
Considering viscous friction that varies spatially and temporally,  the general expressions for entropy production,    free energy,  and entropy extraction rates are derived to a Brownian particle that walks in an overdamped and underdamped media. Via the well known  stochastic approaches to  underdamped and overdamped media,
the thermodynamic expressions first derived at trajectory level then generalized to an ensemble level. To study the non-equilibrium thermodynamic features of a Brownian particle that hops in a medium where its viscosity varies on time,   a  Brownian particle that walks on a periodic isothermal medium (in the presence or absence of load) is considered.  The exact analytical results depict that in the absence of load $f=0$,  the entropy production rate ${\dot e}_{p}$ approaches the entropy extraction rate  ${\dot h}_{d}=0$. This is reasonable since any system which is in contact with a uniform temperature should obey the detail balance condition in a long time limit.    In the presence of load and when the viscous friction decreases either spatially or temporally,   the entropy $S(t)$ monotonously increases with time and saturates to a constant value as $t$ further steps up. The entropy production rate ${\dot e}_{p}$ decreases in time and at steady state (in the presence of load), ${\dot e}_{p}={\dot h}_{d}>0$.  On the contrary,  when the viscous friction increases either spatially or temporally,  the rate of entropy production as well as the rate of entropy extraction monotonously steps up showing that such systems are inherently irreversible. Furthermore, considering a  spatially varying viscosity,  the non-equilibrium thermodynamic features of a Brownian particle that hops in a ratchet potential with load is explored.   In this case, the direction of the particle velocity is dictated by the magnitude of the external load of $f$. Far from the stall load, 
${\dot e}_{p}={\dot h}_{d}>0$ and at stall force    ${\dot e}_{p}={\dot h}_{d}=0$ revealing the system is reversible at this particular choice of parameter.  In the absence of load, ${\dot e}_{p}={\dot h}_{d}>0$ as long as a  distinct temperature difference is retained between the hot and cold baths. Moreover,   considering a multiplicative noise, we explore the thermodynamic features of the model system.

\end{abstract}
\pacs{Valid PACS appear here}% PACS, the Physics and Astronomy
\maketitle

%\section{Introduction}

%%%

 \section{Introduction} 
Understanding the physics of systems out of equilibrium  is challenging  since unlike  equilibrium thermodynamics, non-equilibrium thermodynamics  deals with inhomogeneous  systems  where   the systems  thermodynamic relation complicatedly  relies on the reaction rates. Recently, employing Boltzmann-Gibbs nonequilibrium entropy
along with the entropy balance equation,   the thermodynamic relations of systems which are far from equilibrium  were explored \cite{mu1,mu2,mu3,mu4, mu5,mu6,mu7,mu8,mu9,mu10,mu11,mu12,ta1,mu13,mu14,mu15,mu16}. The  exactly solvable  models   presented in the works  \cite{mu17,muu17},  not only exposed   the factors that affect the entropy production and extraction rates  for a Brownian particle that walks on a discrete lattice system but also uncovered how the  free energy, entropy production, and    entropy extraction rates behave in time. Furthermore,  considering systems that  operate in the quantum realm, the dependence of  thermodynamic  relations   on  the system  parameters  is explored  in the works \cite{mu25,mu26,mu27}. All of these studies  are  vital  to  comprehend    the thermodynamic properties of  systems such as intracellular transport of kinesin or dynein inside the  cell, see  for example the recent works by  
T. Bameta $et$. $al$. \cite{mu28}, D. Oriola $et$. $al$.  \cite{mu29} and  O. Campas $et$. $al$.\cite{mu30}.

More recently     the  general expressions for the free energy, entropy production, and entropy extraction rates to a
 Brownian particle that walks in an overdamped medium was derived  \cite{muuu17}. Furthermore,  considering a Brownian particle that walks in an underdamped medium, the dependence  for   entropy production,    free energy,  and    entropy extraction rates  on the system parameters  was studied \cite{muuu177}.  The results  obtained  by these two works  indicate that as long as the system is  driven out of equilibrium, it constantly produces entropy at the same time it extracts  entropy out of the system.  At steady state,  the rate of entropy production ${\dot e}_{p}$ balances the rate of entropy extraction ${\dot h}_{d}$. At equilibrium, both entropy production and extraction rates become zero. Moreover, the 
entropy production and  entropy extraction rates  are also sensitive to time. As time progresses,  both entropy production and extraction rates increase in time and saturate to  constant values. 

In  the present  study,    we consider a   simple model    where the single  particle  and  its trajectory are considered to be the system as contrasted with the underdamped medium which provides friction and acts as a heat bath. Employing  Boltzmann-Gibbs nonequilibrium entropy, the dependence  for the free energy, entropy production  ${\dot e}_{p}$, and   entropy extraction rates  ${\dot h}_{d}$ on the system parameter  is explored  to  a Brownian particle moving in underdamped and overdamped media. First,  the role of viscous friction is  explored   by considering a viscous friction that varies spatially and temporally. Earlier, Seifert. $et.$ $al.$ \cite{mu6}  introduced  a way of calculating the  entropy production  and extraction rates  at the ensemble level  by first analyzing the thermodynamic relation  at  trajectory level  for a Brownian particle that operates in an overdamped medium.  The alternative approach  by Ge. $et$. $al$. \cite{mar2} and Lee. $et$. $al$. \cite{mar1}  also indicate that under  time reversal operation, the total entropy production and extraction rates can be retrieved. In this work, 
extending these well known  stochastic approaches to  underdamped and overdamped media,  the general expressions for different thermodynamics   relations   are derived. Unlike the previous studies,   we show that  the  entropy production and extraction rates might not saturate  to   a constant value as  the viscous friction   dictates the dynamics of  the system.  Furthermore,  to  both underdamped and overdamped cases,    the rate of entropy production as well as the rate of entropy extraction becomes zero in the long time limit when   the detailed balance  conditions are satisfied.    On the other hand,     our previous exact analytic work  \cite{muuu177} as well as the result obtained in this work  depicts  that at steady state, the  entropy production and  extraction  rates    to  underdamped  case  quantitatively agree  with the overdamped  case.  This is rather puzzling to the  underdamped case   since  the heat exchange due to particle recrossing is unavoidable as long as  a distinct temperature difference is retained  between the hot and cold heat baths.

Some viscous fluids show a change in viscosity when
time changes. This is because as the fluid shear stress
changes in time, so does the viscosity. Often, the dynamics
of systems with self-organized criticality also can be explored
by considering time dependent diffusion (viscous
friction) and drift terms \cite{marr1,marr2, marr3}. Some studies have also 
focused on calculating the mean first passage time by considering a  time dependant diffusion term \cite{muuuu17}. To explore the non-equilibrium thermodynamic features of a Brownian particle that hops in a medium where its
viscosity depends on time, we consider a Brownian particle
that walks on a periodic isothermal medium (in the presence or absence of load).
  The exact analytical results depicts  that  in the absence of  load $f=0$,  ${\dot e}_{p} = {\dot h}_{d}=0$. This  is reasonable since any  system which is in contact with a uniform  temperature should obey the detail balance condition only in a long time limit. This can be intuitively  comprehended  on physical grounds.  When the particle operates at a finite time, the system operates irreversibility and in this regime,  the second law of thermodynamics  states that the change in entropy $\Delta S(t)>0$. As one  can see    that if the thermodynamic quantities are evaluated   in the time interval between $t=0$ and any time $t$, always the change in entropy,  entropy production, and entropy extraction rates become greater than zero revealing such systems are inherently  irreversible. Moreover, we show that when a distinct temperature difference is not retained  between the hot and cold  baths, in  absence of load,  ${\dot e}_{p}={\dot h}_{d}=0$ showing that the system is reversible.  In the presence of load and when the viscous friction  decreases  in time,  we show that the entropy $S(t)$ monotonously increases   with time and saturates to  a constant value as $t$ further steps up.  The entropy production rate ${\dot e}_{p}$ decreases in time and at steady state (in the presence of load), ${\dot e}_{p}={\dot h}_{d}>0$ which agrees with the results shown in the works \cite{muuu177}.  On the contrary,  when the  viscous friction  increases in time,  the rate of entropy production as well as the rate of entropy extraction monotonously    steps up   showing that such systems are  inherently irreversible.

Most of the previous studies have  also focused  on exploring the thermodynamic feature of  systems such as    Brownian heat engines by assuming  temperature invariance viscous friction. In reality,  the viscous  friction of a  medium tends to decrease as the temperature of the  medium increases. This is because as the intensity of the background temperature increases, the force   of interaction  between neighboring molecules decreases. In this  paper,    considering a  spatially varying  viscosity,  the non-equilibrium thermodynamic features of   a Brownian particle that  hops in a ratchet potential with load  is explored.   The potential  is also coupled with a spatially  varying temperature. In this case, the direction of the particle velocity is dictated by the magnitude of the external load of $f$. As one can note that the steady state velocity of the engine is positive   when $f$ is smaller and the engine acts as a heat engine. In this regime ${\dot e}_{p}={\dot h}_{d}>0$.  When $f$ steps  up, the velocity of the particle steps down, and at stall force, we find that ${\dot e}_{p}={\dot h}_{d}=0$ revealing  that the system is reversible  at this particular choice of parameter.  For large force, the current is negative and the engine acts as a refrigerator. In this region ${\dot e}_{p}={\dot h}_{d}>0$. In the  absence of load, ${\dot e}_{p}={\dot h}_{d}>0$ as long as a  distinct temperature difference is retained between the hot and cold baths.

At this point, we want also to stress that  most of the previous works have focused on calculating
the thermodynamic features of different model systems
by considering additive noise. On the contrary,  most realistic
 systems  such as neuron  system   can be also described by Langevin
equations with multiplicative noise were in this case,  the noise
amplitude varies spatially \cite{mar12}.  In this paper,  we
study how thermodynamic  features of such systems  behave.

The rest of the paper is organized as follows: in Section II, we  derive the expression for  various thermodynamic  relations to a Brownian particle walking in overdamped and underdamped  media.  In Section III, the role of viscous friction  is studied by considering viscous friction that varies spatially and temporally. In section IV, we explore the model system in the presence of multiplicative noise. Section V deals with a summary and conclusion.

\section{Free energy, entropy production and entropy extraction rates }  
Recently,    the dependence  for   entropy production,    free energy,  and    entropy extraction rates on  the system parameters  was explored   \cite{muuu177}  by  considering  a Brownian particle that walks in a medium where its viscous friction is insensitive to time or position. However, in most realistic  systems,  the   viscous friction of the  medium  varies  spatially or  temporally.  To address this issue,  let us  consider  a Brownian  particle that   moves  in an underdamped medium along the   potential  $U(x)=U_{s}(x)+fx$ where  $U_{s}(x)$ and $f$ are the periodic  potential  and the external force, respectively. Next, the      relation  for    the entropy production and  extraction rates   will be derived considering a spatially varying  viscous friction.

\subsection{Underdamped case}
{\it Derivation for entropy production and entropy extraction rates.\textemdash} Let us  consider a single  Brownian particle that is arranged  to undergo a random walk in an underdamped medium. Here the single-particle and
its trajectory are considered to be the system as contrasted with the underdamped medium which provides friction and acts as a heat bath.   The dynamics  of the system is governed  by 
\begin{equation}
m{dv\over dt} = -\gamma (x,t){dx\over dt}+ {d U(x) \over dx}  + \sqrt{2k_{B}\gamma (x,t) T(x)}\xi(t).
\end{equation}
For simplicity,  Boltzmann constant  $k_{B}$  is assumed to be unity. The random noise $\xi(t)$ is assumed to be Gaussian white noise satisfying the relations 
$\left\langle  \xi(t) \right\rangle =0$ and $\left\langle \xi(t)  \xi(t') \right\rangle=\delta(t-t')$.  The viscous  friction  $\gamma (x,t)$  and  $T(x)$  are  assumed  to vary spatially along the medium.

For underdamped case,  the   Fokker-Plank equation   is given by
\begin{eqnarray}
{\partial P\over \partial t}&=&-{\partial (vP) \over \partial x}-{1 \over m}{\partial(U'(x)P) \over \partial v}+ \nonumber \\
&&{\gamma(x,t) \over m}{\partial (vP) \over \partial v}+{\gamma(x,t) T(x) \over m^2}{\partial^2 P \over \partial v^2}
\end{eqnarray}
where $P(x,v,t)$ is the probability  of finding the particle at particular position $x$, velocity $v$ and time $t$.

%For overdamped case,
 %as discussed   by Sancho. $et$ .$at$  \cite{am3}  and Jayannavar   $et$ .$at$  \cite{am33, 
%the above Langevin equation (1)    converges to 
%textcolor {red}{\begin{eqnarray}
%gamma(x,t){dx\over dt}&=&{-\partial U(x)\over \partial x} -{1\over 2 %gamma(x,t)}{\partial\over   \partial x}(\gamma(x,t)T(x))+ \nonumber \\
%&&\sqrt{2k_{B}\gamma(x,t) T(x)}\xi(t)
%\end{eqnarray}}
  %which corresponds  to Stratonovich interpretation \cite{am1,am2}. 
 
For convenience, Eq. (2)  can be rearranged     as 
\begin{eqnarray}
{\partial P\over \partial t}&=& -(k+{\partial J' \over \partial v} )
\end{eqnarray}
where 
\begin{eqnarray}
k=v{\partial P \over \partial x} ={\partial J \over \partial x}
\end{eqnarray}
and 
\begin{eqnarray}
J'= -{\gamma (x,t)\over m}vP+{1\over m}(U'P) -{\gamma (x,t) T(x)\over m^2}{\partial P\over \partial v}.
\end{eqnarray}
From Eqs. (4) and (5), one gets 
\begin{eqnarray}
{\partial P \over \partial v}= -{m^{2}J'\over \gamma (x,t)T(x)}+{mU'P\over \gamma (x,t)  T(x)}-{m vP\over T(x)}
\end{eqnarray}
and  
\begin{eqnarray}
{\partial P \over \partial x}= {k\over v}.
\end{eqnarray}
Next we  derive  the expressions for the  entropy production by considering two cases.

{\it Case1.\textemdash}   Here we want to stress  that the  approach  by Lee. $et$. $al$. \cite{mar1} and Ge. $et$. $al$. \cite{mar2} indicate that under  time reversal operation, the total entropy production and extraction rate  can  be obtained.  Particularly, the analysis by Ge. $et$. $al$. \cite{mar2} shows that  the entropy production rate   is given by
\begin{eqnarray}
{\dot e}_{p}&=&-\int {1\over T(x) \gamma(x,t)}\left[F-{T(x) \gamma(x,t)\nabla_{v}\ln(P)\over m}\right]^2P dx dv \nonumber \\
&=&-\int {1\over T(x) \gamma(x,t)}\left[F-{T(x) \gamma(x,t){\partial P\over \partial v}\over m P}\right]^2P  dx dv
\end{eqnarray}
where $F=-T(x) \gamma(x,t)+U'$.
Substituting Eq. (6) into Eq. (8), one gets 
\begin{eqnarray}
{\dot e}_{p}&=& \int {m^{2}J'^{2} \over P T(x) \gamma(x,t)} dx dv 
\end{eqnarray}
On the other hand, the entropy  extraction rate can be  found   via the method devloped by Ge. $et$. $al$. \cite{mar2} as
\begin{eqnarray}
{\dot h}_{d}&=&-\int {1\over T(x) }\left[T(x) \gamma(x,t)+{T(x) \gamma(x,t)\nabla_{v}\ln(P)\over m}\right]P v dxdv \nonumber \\
&=&-\int {1\over T(x) }\left[T(x) \gamma(x,t)+{T(x) \gamma(x,t){\partial P\over \partial v}\over m P}\right]P v dxdv.
\end{eqnarray}
Substituting Eq. (6) into Eq. (10)  leads to
 \begin{eqnarray}
{\dot h}_{d} &=& \int  {(U'J-vmJ') \over T(x) } dx dv.
\end{eqnarray}  Here   ${\dot e}_{p}={d e_{p}\over dt} $ and   ${\dot h}_{d}={d h_{d}\over dt}$  denote the entropy production   and  extraction rates.
The  above expressions  (for  the entropy production and extraction rates) can be  also derived at the ensemble level  via the  approach stated in the work \cite{mu7}. 

{\it Case2.\textemdash}  One can also rederive  the expressions for the  entropy production  and extraction rates  at the ensemble level by first analyzing the entropy  of the system  at the trajectory level as
\begin{eqnarray}
s(t)=-ln P(x,v,t)
\end{eqnarray}
where $x(t)$ denotes the  stochastic trajectory.  
The rate of entropy
change at trajectory level is then  given by
\begin{eqnarray}
{\dot s}(t)=-{\partial_{t} P(x,v,t) \over P(x,v,t)}-{\partial_{x} P(x,v,t) \over P(x,v,t)}{\dot x}-{\partial_{v} P(x,v,t) \over P(x,v,t)}{\dot v}.
\end{eqnarray}

Substituting Eqs . (6) and (7) into Eq. (13), the entropy production and dissipation
rates at trajectory level are given as
\begin{eqnarray}
{\dot e}_{p}^{*}=-{\partial_{t} P(x,v,t) \over P(x,v,t)}+{m^{2}J'\over \gamma (x,t)T(x)P(x,v,t)}{\dot v}-{k \over P(x,v,t)}
\end{eqnarray}
and 
\begin{eqnarray}
{\dot h}_{d}^{*}={mU'\over \gamma (x,t) T(x)}{\dot v}-{m\gamma(x,t) v\over T(x)}{\dot v}.
\end{eqnarray}
Because averaging overall trajectories yields $\left\langle {\dot v}|x\right\rangle={J'\over P(x,t,v)}$ and   $\int \partial_{t} P(x,v,t ) =0$,    after some algebra one gets 
\begin{eqnarray}
{\dot e}_{p}&=& \int {m^{2}J'^{2} \over P T(x) \gamma(x,t)} dx dv -\int P v dv
\end{eqnarray}
and 
\begin{eqnarray}
{\dot h}_{d} &=& \int  {m(U'-v\gamma (x,t))J' \over T(x)\gamma (x,t) } dx dv,
\end{eqnarray} respectively.
When a periodic boundary condtion is imposed,   Eqs. (9) and (16) as well as Eqs. (11) and  (17) converge to
\begin{eqnarray}
{\dot e}_{p}&=& \int {m^{2}J'^{2} \over P T(x) \gamma(x,t)} dx dv 
\end{eqnarray}
and 
\begin{eqnarray}
{\dot h}_{d} &=& -\int  {vmJ' \over T(x) } dx dv.
\end{eqnarray} 

The heat dissipation rate ${\dot H}_{d}$    can  be calculated  \cite{am4,am5} as 
\begin{eqnarray}
{\dot H}_{d}  
&=&-\left\langle \left(-\gamma(x,t){\dot x}+ \sqrt{2k_{B}\gamma(x,t) T(x)}\right).{\dot x}\right\rangle  \nonumber \\
&=&-\left\langle m{vdv\over dt}  +v U'(x)   \right\rangle.
\end{eqnarray}
Our previous  analysis also  suggests  \cite{mu17,muu17,muuu17} that    the entropy extraction rate ${\dot h}_{d}$  can be  expressed as 
\begin{eqnarray}
{\dot h}_{d}  
&=&-\int \left({m{vdv\over dt}  +v U'(x) \over T(x)} \right)P dxdv.
\end{eqnarray}
 One should note that Eq. (21) is exact and does not depend on any boundary condition. 
Since ${d S(t)\over dt}$ and ${\dot h}_{d}  $ are computable,  the entropy production rate  can be readily  obtained as
\begin{eqnarray}
{\dot e}_{p}&=&{d S(t)\over dt}+{\dot h}_{d}. 
\end{eqnarray}
In the long time limit,  ${d S(t)\over dt}=0$ which implies ${\dot e}_{p}={\dot h}_{d}>0$ at steady state  and ${\dot e}_{p}={\dot h}_{d}=0$  at stationary state.

%At steady state  which implies that  ${\dot e}_{p}={\dot h}_{d}>0$. For isothermal case, at (approaching equilibrium),  

%Moreover, for the case where the probability distribution is  either periodic or  vanishes  at the boundary, Tome $et$. $at.$ \cite{ta1} derived the expressions for the entropy production and entropy extraction rates for isothermal case.  
%The expression $k$ vanishes after imposing a boundary condition.

%In the next sections  we  show that indeed  Eqs. (8)  and  (16)  as well Eqs. (9) and (15) agree  as long as a periodic boundary condition is imposed.  
Once  the  expressions for 
${\dot S}(t)$, ${\dot e}_{p}(t)$  and ${\dot h}_{d}(t)$  are  computed as a function of   time $t$, the analytic expressions for  the change in entropy production,  heat dissipation  and  total entropy can be found analytically via   
$
\Delta h_d(t)= \int_{0}^{t}{\dot h}_{d}(t)dt $,
$\Delta e_{p}(t)= \int_{0}^{t}  {\dot e}_{p}(t)  dt $ and 
$\Delta S(t) =\int_{0}^{t} {\dot S}(t)dt $
where $\Delta S(t)=\Delta e_p(t)-\Delta h_d(t)$.

{\it Derivation for free energy.\textemdash}  Our next objective is to write the expression for  the free energy  in terms of  ${\dot E}_{p}(t)$   and  ${\dot H}_{d}(t)$ where  ${\dot E}_{p}(t)$   and  ${\dot H}_{d}(t)$ are the terms that are associated with ${\dot e}_{p}(t)$   and  ${\dot h}_{d}(t)$.    As discussed  before, the heat dissipation rate   is either  given by Eq. (20) (for any cases) or if a periodic boundary condition is imposed, ${\dot H}_{d}(t)$ is given by 
\begin{eqnarray}
{\dot H}_{d} &=&-\int  m(v)J'   dx dv
\end{eqnarray}
which is   notably different from Eq. (19), due to the  term $T(x)$. 
The term  associated  to ${\dot e}_{p}$  is given by 
\begin{eqnarray}
{\dot E}_{p}&=& -\int {m^{2} J'^{2} \over P \gamma(x,t)} dx dv.
\end{eqnarray}
The    entropy balance  equation 
\begin{eqnarray}
{d S^T(t)\over dt}&=&{\dot E}_{p}-{\dot H}_{d}
\end{eqnarray}
is associated to Eqs. (11) or (17)  except the term $T(x)$.
Once again,   employing  the  expressions for 
${\dot S}^T(t)$, ${\dot E}_{p}(t)$  and ${\dot H}_{d}(t)$,  one can get    
$\Delta H_d(t)= \int_{0}^{t}{\dot H}_{d}(t)dt$, 
$\Delta E_{p}(t)= \int_{0}^{t} {\dot E}_{p}(t) dt $ and 
$\Delta S(t)^T =\int_{0}^{t} {\dot S}(t)^{T}dt$
where $\Delta S(t)^T=\Delta E_p(t)-\Delta H_d(t)$.

On the other hand,  the  expression for the internal energy  has a form 
\begin{eqnarray}
{\dot E}_{in} = \int ({\dot K}+ v U'_{s}(x))P(x,v,t)dvdx
\end{eqnarray}
where   ${\dot K}=m{vdv\over dt}$  and $U'_{s}$ denote the rate of kinetic and   potential energy, respectively. The  network work done by the system  
\begin{eqnarray}
{\dot W}&=& \int v f P(x,v,t)dvdx
\end{eqnarray}
explicitly   depends on the velocity $V$ and the load $f$. 
In terms of ${\dot H}$ and ${\dot W}$, the rate of  the internal energy  is given  by 
\begin{eqnarray}
{\dot E}_{in} = -{\dot H}_{d}(t)-{\dot W}
\end{eqnarray}
and after some algebra, the first law of thermodynamics  can be written as  
\begin{eqnarray}
\Delta E_{in}= -\int_{0}^{t}( {\dot H}_{d}(t)+{\dot  W}) dt.
\end {eqnarray}

Rearranging  some terms, one gets 
the rate of free energy  as  ${\dot F}={\dot E}-T{\dot S}$ for isothermal case and  ${\dot F}={\dot E}-{\dot S}^T$ for nonisothermal case  where ${\dot S}^T={\dot E}_{p}-{\dot H}_{d}$. The rate of free energy dissipation 
\begin{eqnarray}
{\dot F}&=&{\dot E}_{in}- {\dot S}^T \nonumber \\
&=&{\dot E}_{in}-{\dot E}_{p}+{\dot H}_{d}
\end{eqnarray}
can be  expressed as  a definite  integral as
\begin{eqnarray}
\Delta F(t)&=&-\int_{0}^{t} \left(  {\dot W}+ {\dot E}_{p}(t)   \right)dt.
\end{eqnarray} 
For isothermal case, at  quasistatic limit where the velocity  approaches  zero  $ v=0$, ${\dot E}_{p}(t) =0$ and ${\dot H}_{d}(t) =0$ and far from quasistatic limit 
$E_{p}={\dot H}_{d}>0$  which is  expected as   the particle  operates irreversibly. 

\subsection{Overdamped case}

{\it Derivation for free energy,  entropy production and  entropy extraction rates.\textemdash} For overdamped case,
 as discussed   by Sancho. $et$ .$at$  \cite{am3}  and Jayannavar   $et$ .$at$  \cite{am33}, 
Eq.  (1)    converges   to     
\begin{eqnarray}
\gamma(x,t){dx\over dt}&=&{-\partial U(x)\over \partial x} -{(\gamma'(x,t)T(x)+\gamma(x,t)T'(x))\over 2 \gamma(x,t)}+\nonumber \\
&&\sqrt{2k_{B}\gamma(x,t) T(x)}\xi(t)
\end{eqnarray}
which corresponds  to  the Stratonovich interpretation \cite{am1,am2}.
 The corresponding Fokker Planck equation is given by 
\begin{eqnarray}
{\partial P(x,t)\over \partial t}&=&{\partial  \over \partial x}\left({U'(x)\over \gamma(x,t)}+{(\gamma'(x,t)T(x)+\gamma(x,t)T'(x))\over 2 \gamma(x,t)^2}\right)P(x,t)+\nonumber \\
&&{\partial  \over \partial x}\left({T(x)\over \gamma(x,t)}{\partial P(x,t)\over \partial x}\right)
\end{eqnarray}
which can be rewritten as 
\begin{eqnarray}
{\partial P(x,t)\over \partial t}&=&-{\partial J \over \partial x}
\end{eqnarray}
where
\begin{eqnarray}
J&=&-\left({U'(x)P(x,t)\over  \gamma(x,t)}+{P(x,t)(\gamma'(x,t)T(x)+\gamma(x,t)T'(x))\over 2 \gamma(x,t)^2}\right)-\nonumber \\
&&\left({T(x)\over \gamma(x,t)}{\partial P(x,t)\over \partial x}\right).
\end{eqnarray}

The rate of entropy
change at trajectory level is given by
\begin{eqnarray}
{\dot s}(t)=-{\partial_{t} P(x,t) \over P(x,t)}-{\partial_{x} P(x,t) \over P(x,t)}{\dot x}.
\end{eqnarray}
On the other hand,  from Eq. (35) one gets 
\begin{eqnarray}
{\partial P \over \partial x}= -{\gamma (x,t) J\over T(x)}-{U'(x)P(x,t)\over  T(x)}-{P(x,t)(\gamma'(x,t)T(x)+\gamma(x,t)T'(x))\over 2 \gamma(x,t)T(x)}.
\end{eqnarray}

Substituting Eq. (37) into Eq. (36), the entropy production and dissipation
rates at trajectory level are given as
\begin{eqnarray}
{\dot e}_{p}^{*}=-{\partial_{t} P(x,t) \over P(x,t)}+{\gamma (x,t)J\over T(x)P(x,t)}{\dot v}
\end{eqnarray}
and 
 \begin{eqnarray}
{\dot h}_{d}^{*}={U'(x)\over  T(x)}+{(\gamma'(x,t)T(x)+\gamma(x,t)T'(x))\over 2 \gamma(x,t)T(x)}.
\end{eqnarray}
Because averaging overall trajectories yields $\left\langle {\dot x}|x\right\rangle={J\over P(x,t)}$ and   $\int \partial_{t} P(x,t ) =0$, after some algebra one gets 
\begin{eqnarray}
{\dot e}_{p}&=& \int {\gamma(x,t) J^{2} \over P T(x)} dx 
\end{eqnarray}
and 
\begin{eqnarray}
{\dot h}_{d} &=&\int\left({JU'(x)\over  T(x)}+{J(\gamma'(x,t)T(x)+\gamma(x,t)T'(x))\over 2 \gamma(x,t)T(x)}\right) dx 
\end{eqnarray} respectively. One should note that Eqs. (18) and (19) (underdamped case)  as well as Eqs. (40) and (41) (overdamped case) approach  
\begin{eqnarray}
{\dot h}_{d}  ={\dot e}_{p}
&=&\int \left({J U'(x) \over T(x)} \right)dx
\end{eqnarray}
at steady state ($v {dv\over dt}=0$), 
as long as a 
periodic boundary condition is imposed.
From Eqs. (40) and (41), it is evident that  when detailed balance conditions are satisfied  the velocity of  equivalently the current $J=0$ and as a result $ {\dot e}_{p}=0$ and $ {\dot h}_{d}=0$. Far from equilibrium, $J>0$, and in this case when the  viscous  friction decreases either spatially or  temporally, ${\dot e}_{p}$ and ${\dot h}_{d}$  approach to a constant value.  When the system is driven out of equilibrium and when  viscous friction  increases spatially and temporally, $ {\dot e}_{p}$ and $ {\dot h}_{d}$ 
monotonously  step up.
Moreover, from Eq. (41),  the heat dissipation rate   is  derived as 
\begin{eqnarray}
{\dot H}_{d} &=& \int \left(JU'(x)+{J(\gamma'(x,t)T(x)+\gamma(x,t)T'(x))\over 2 \gamma(x,t)} \right) dx .
\end{eqnarray}
On the other hand, the term  ${\dot E}_{p}$ is  related to  ${\dot e}_{p}$ and it is given by 
\begin{eqnarray}
{\dot E}_{p}&=& \int {\gamma(x,t) J^{2} \over P } dx.
\end{eqnarray}
The new entropy balance equation has a simple form   
\begin{eqnarray}
{d S^T(t)\over dt}&=&{\dot E}_{p}-{\dot H}_{d}.
\end{eqnarray}
Furthermore, the internal energy 
\begin{eqnarray}
{\dot E}_{in} = \int J U'_{s}(x) dx
\end{eqnarray}
has functional dependence on the current $J$ and the  potential profile $U_s$. 
The total work done is then given by 
\begin{eqnarray}
{\dot W}&=&  \int \left(Jf+{J(\gamma'(x,t)T(x))\over 2 \gamma(x,t)} +{JT'(x)\over 2 }\right)dx.
\end{eqnarray}
The first law of thermodynamics  can be written as  
\begin{eqnarray}
{\dot E}_{in} = -{\dot H}_{d}(t)-{\dot W}.
\end{eqnarray}
 The change in the internal energy   reduces to  
$
\Delta E_{in}= -\int_{0}^{t}( {\dot H}_{d}(t)+{\dot  W}) dt.
$
Once again 
the rate of free energy dissipation  can be written as 
${\dot F}={\dot E}_{in}- {\dot S}^T 
={\dot E}_{in}-{\dot E}_{p}+{\dot H}_{d}$. The change in the free energy is then  given by  
$\Delta F(t)=-\int_{0}^{t} \left(  {\dot W}+ {\dot E}_{p}(t)   \right)dt.
$

\section{  Time dependent    viscous friction  }

Some viscous fluids show a change in viscosity when
time changes. This is because as the fluid shear stress
changes in time, so does the viscosity. Often, the dynamics
of systems with self-organized criticality also can be explored
by considering time dependent diffusion (viscous
friction) and drift terms \cite{marr1,marr2, marr3}. Some studies have  also
focused on calculating the mean first passage time by considering a  time dependant diffusion term \cite{muuuu17}. To explore the non-equilibrium thermodynamic features of a Brownian particle that hops in a medium where its
viscosity depends on time, we consider a Brownian particle
that walks on a periodic isothermal medium (in the
presence or absence of load) where its viscosity is given
by 
\begin{eqnarray}
\gamma(t)&=&{1\over g(1+t^z)}.
\end{eqnarray}
In this case,   the corresponding  Fokker Planck equation  in  overdamped medium is given as 
\begin{eqnarray}
{\partial P(x,t)\over \partial t}&=&{\partial  \over \partial x}\left({f\over \gamma(t)}\right)P(x,t)+\nonumber \\
&&{\partial  \over \partial x}\left({T\over \gamma(t)}{\partial P(x,t)\over \partial x}\right).
\end{eqnarray}
Imposing a periodic boundary condition $P(0,t)=P(L,t)$  and  let us  choose a  Fourier cosine series  
\begin{eqnarray}
P(x,t)=\sum_{n=0}^\infty b_{n}(t)cos\left({n\pi \over L_0}(x+{f\over \gamma(t)})\right)
\end{eqnarray}
as a possible solution. 
After some algebra,  we get  
 the   probability distribution  as 
\begin{eqnarray}
P(x,t)&=&\sum_{n=0}^\infty \cos\left[{n\pi \over L_0}\left(x+f(gt+{gt^{z+1}\over(z+1)})\right)\right]\zeta.
\end{eqnarray}
where
\begin{eqnarray}
\zeta&=&e^{-{(n\pi )^2T\left(gt+{gt^{z+1}\over(z+1)}\right) \over L^2}}.
\end{eqnarray}
Here $f$ is the external load and $T$ is the temperature of the medium.
The current is then given by 
\begin{eqnarray}
J(x,t)&=&-\left[{f P(x,t)\over \gamma(t)} + {T\over \gamma(t)}{\partial P(x,t) \over \partial x}\right].
\end{eqnarray}
The current $J(x,t)>0$, only when $f \ne 0$ since $\gamma(x)$ is not the necessary parameter to keep the system out of equilibrium. 
 As stated before, $
{\dot e}_{p}= {\dot h}_{d}+{d S(t)\over dt}$ where 
$
{d S(t)\over dt}=-\int {J\over P(x,t)}{\partial \over \partial x}  P(x,t)  dx$.  
After some algebra, we write  
\begin{eqnarray}
{d S(t)\over dt}&=&-\int J {\sum_{n=0}^\infty {{n\pi\over 2} \cos\left[{n\pi \over L_0}\left(x+f(gt+{gt^{z+1}\over(z+1)})\right) \right] \zeta \over \sum_{n=0}^\infty \cos\left[{n\pi \over L_0}\left(x+f(gt+{gt^{z+1}\over(z+1)})\right)\right]\zeta}} dx.
\end{eqnarray}
 The entropy production   and  entropy extraction rates   are given by the relations   
\begin{eqnarray}
{\dot e}_{p}&=& \int {J^{2} \over P(x,t) T g(1+t^z) } dx 
\end{eqnarray}
and 
\begin{eqnarray}
{\dot h}_{d} &=&\int\left({Jf\over  T}\right) dx.
\end{eqnarray} 
Substituting Eqs. (52) and (54) into Eqs. (56) and (57), one can explore how ${\dot e}_{p}$ and ${\dot h}_{d}$ depend on time. For $z \ge 0$, ${\dot e}_{p}$ and ${\dot h}_{d}$  decrease and approach to a constant  value at steady state. When $z<0$, ${\dot e}_{p}$ and ${\dot h}_{d}$ step up continuously. 
Hereafter, for simplicity, the parameter $g$ is considered to be  a constant. 
 Furthermore, the heat dissipation rate   is given by 
\begin{eqnarray}
{\dot H}_{d} &=& \int \left(Jf \right) dx 
\end{eqnarray}
while the term  ${\dot E}_{p}$ is  given by 
\begin{eqnarray}
{\dot E}_{p}&=& \int {J^{2} \over P(x,t)g(1+t^z) } dx.
\end{eqnarray}
On the other hand, the internal energy has a form  
\begin{eqnarray}
{\dot E}_{in} = \int J U'_{s}(x) dx.
\end{eqnarray}
The total work done is then given by 
\begin{eqnarray}
{\dot W}&=&  \int \left(Jf \right)dx.
\end{eqnarray}
The first law of thermodynamics  can be written as  
\begin{eqnarray}
{\dot E}_{in} = -{\dot H}_{d}(t)-{\dot W}
\end{eqnarray}
Hereafter, whenever we plot any figures,  we use    the following dimensionless 
load ${\bar f}=fL_{0}/T_{c}$, ${\bar U}=U/T_{c}$, temperature ${\bar \tau}=T(x) /T_{c}$ where $T_c$ is the reference temperature.
We  also introduced  dimensionless parameter  ${\bar x}=x/L_{0}$, ${\bar v}=vm/ \gamma L_{0}$ and  ${\bar t}=t \gamma /m$.  Hereafter the bar will be dropped.
From now on all  the  figures will be plotted in terms of the dimensionless parameters.

The expression for the rate of entropy production as well as entropy extraction rate  can be  readily calculated via Eqs. (56) and (57). In the absence of  load $f=0$,  ${\dot e}_{p} = {\dot h}_{d}=0$. This  is reasonable since any  system which is in contact with a uniform  temperature should obey the detail balance condition in a long time limit. When a distinct temperature difference is retained  between the hot and cold  baths, in  absence of load,  ${\dot e}_{p}={\dot h}_{d}=0$ showing that such a system is inherently  reversible.  In the presence of load and when the viscous friction increases in time (see Fig. 1),  ${\dot e}_{p}$ and ${\dot h}_{d}$  decrease in time and as time further
steps up ${\dot e}_{p}$ and ${\dot h}_{d}$ monotonously increase  in  time as shown in Fig. 1.  Figure 1 is plotted by fixing $\tau=1.0$,  $f=1.0$ and $z=-0.5$. On the contrary,  in the presence of load and when the viscous friction decreases in time (see Fig. 2),  ${\dot e}_{p}$ and ${\dot h}_{d}$ monotonously decrease with time and saturate to  a constant value as $t$ further steps up.  Fig. 2 is plotted by fixing $\tau=1.0$,  $f=1.0$ and $z=1.0$. 
\begin{figure}[ht]
\centering
%\subfigure[Bild a.] % caption for subfigure a
{
    %\label{fig:sub:a}
    \includegraphics[width=6cm]{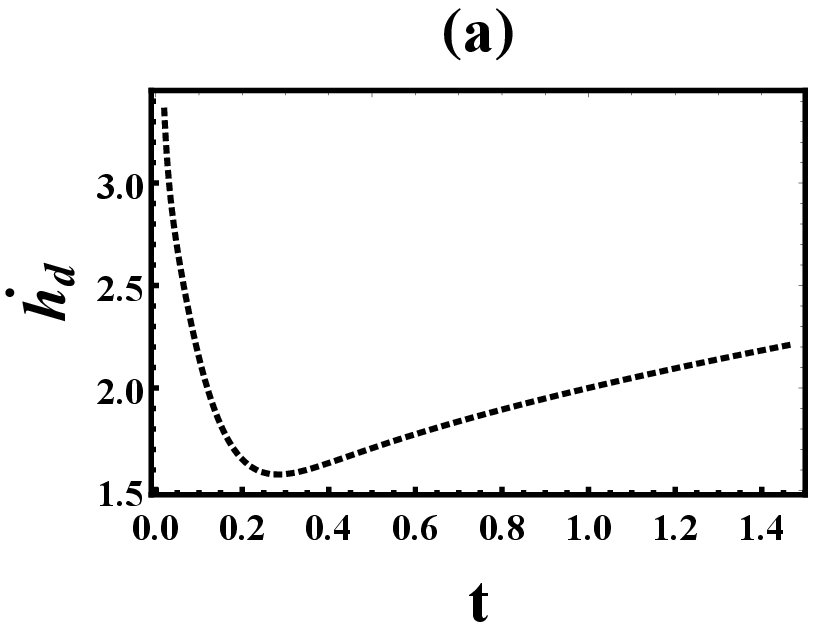}}
\hspace{1cm}
%\subfigure[Bild b.] % caption for subfigure b
{
    %\label{fig:sub:b}
    \includegraphics[width=6cm]{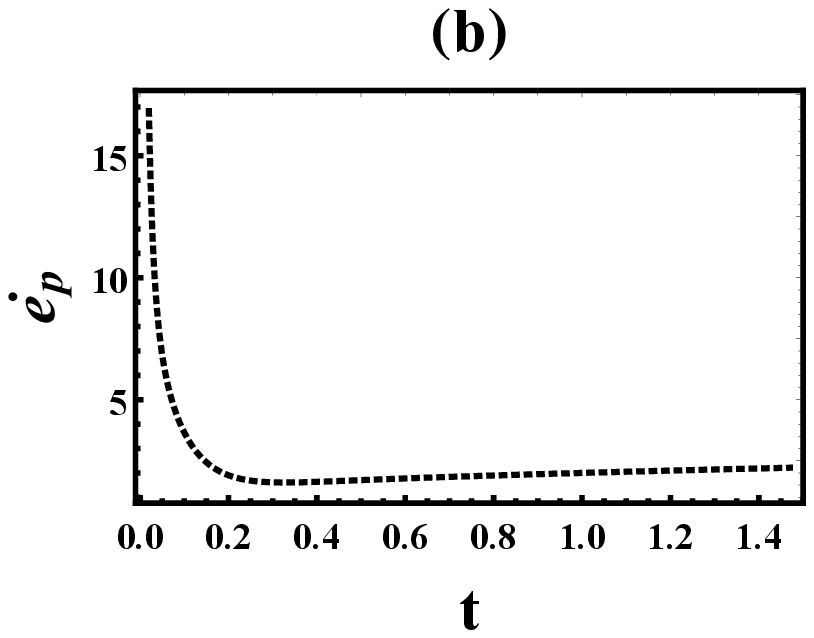}
}
\caption{ (Color online) (a) The entropy extraction rate ${\dot h}_{d}(t)$ as a function of $t$ evaluated analytically  by substituting Eqs. (52) and (54) into Eq. (57). (b) The plot of entropy production rate ${\dot e}_{p}(t)$ as a function of $t$.
${\dot e}_{p}(t)$  is analyzed  analytically by substituting Eqs. (52) and (54) into Eq. (56).  The two  figures exhibit that ${\dot e}_{p}(t)$ and  ${\dot h}_{d}(t)$  decrease in time  and as time further steps  up,  ${\dot e}_{p}(t)$ and  ${\dot h}_{d}(t)$ increase. In the both figures, the parameters are fixed as  $f=1.0$, $\tau=1.0$ and   $z=-0.5$. 
} 
\label{fig:sub} % caption for the whole figure
\end{figure}
\begin{figure}[ht]
\centering
%\subfigure[Bild a.] % caption for subfigure a
{
    %\label{fig:sub:a}
    \includegraphics[width=6cm]{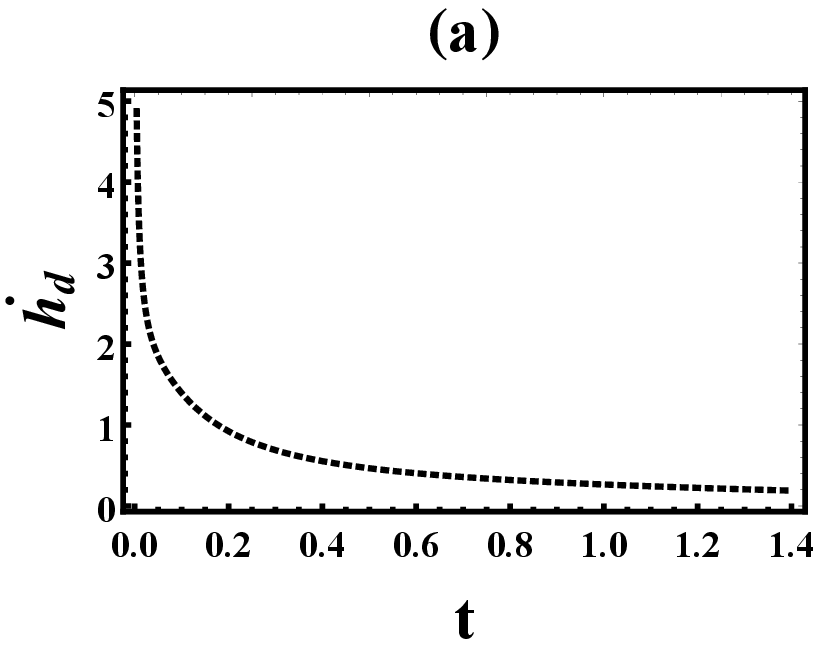}}
\hspace{1cm}
%\subfigure[Bild b.] % caption for subfigure b
{
    %\label{fig:sub:b}
    \includegraphics[width=6cm]{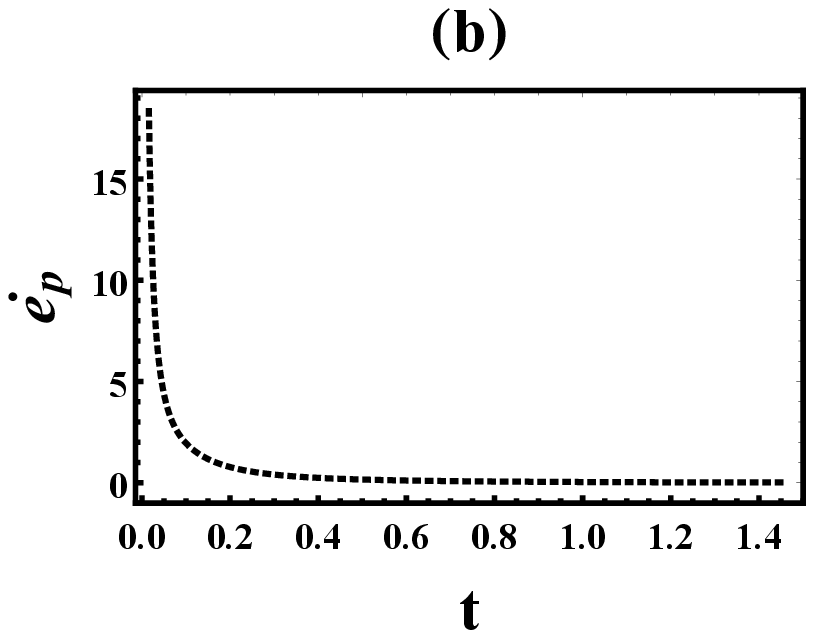}
}
\caption{ (Color online)  (a)  ${\dot h}_{d}(t)$ as a function of $t$ evaluated analytically  by substituting Eqs. (52) and (54) into Eq. (57). (b)  ${\dot e}_{p}(t)$ as a function of $t$.
${\dot e}_{p}(t)$  is analyzed  analytically by substituting Eqs. (52) and (54) into Eq. (56).  The  figure exhibits that ${\dot e}_{p}(t)$ and  ${\dot h}_{d}(t)$  decrease in time  and as time further steps  up,  ${\dot e}_{p}(t)$ and  ${\dot h}_{d}(t)$ approach  a constant value. In the both figures, the parameters are fixed as  $f=1.0$, $\tau=1.0$ and   $z=1.0$. 
} 
\label{fig:sub} % caption for the whole figure
\end{figure}

\section{ Spatially varying   viscous friction }

Most of the previous studies have focused on exploring the thermodynamic feature of systems such as    Brownian heat engines by assuming temperature invariance viscous friction.  However, various studies have indicated that the viscous friction of a  medium tends to decrease as the temperature of the medium increases  \cite{mar15, mar16, aa15}. Particularly in the liquid medium, the viscosity decreases as the intensity of the background temperature steps up. This is because when the temperature of the medium increases,  more molecules start vibrating, and as a result their speed increases. This speedy motion of the molecules creates a reduction in interaction time between neighboring molecules. At the macroscopic level, there will be a reduction in the intermolecular force, and hence reduced viscosity of the fluid. Consequently, when the temperature of the viscous medium decreases, the viscous friction in the medium decreases.

In this  paper,    considering a  spatially varying  viscosity,  the non-equilibrium thermodynamic features of   a Brownian particle that  hops in a ratchet potential with load  is explored.   The potential  is also coupled with a spatially  varying temperature. 

{\it The model .\textemdash} Let us  consider a Brownian particle that walks in a piecewise linear potential with an external load $U(x)=U_{s}(x) + fx$, where the ratchet potential
$U_{s}(x)$ is given by
\begin{equation}
  U_{s}(x)=\left\{
  \begin{array}{ll}
    2U_{0}\left({x\over L_{0}}\right),& \text{if}~~~ 0 \le x \le {L_{0}\over 2};\\ %cr
    2U_{0}\left(1-{x\over L_{0}}\right),& \text{if} ~~~{L_{0}\over 2} \le x \le L_{0}. %\cr
   \end{array}
   \right.
\end{equation}
Here $U_{0}$ and $L_{0}$ denote  the barrier height and the width of the ratchet potential,
respectively. $f$ designates  the external force. 
The potential exhibits its maximum value
$U_{0}$ at  $x={L_{0}\over 2}$ and its minima at $x=0$  and $x={L_{0}}$.
The 
spatially varying temperature is arranged as 
\begin{equation}
T(x)=\left\{
\begin{array}{ll}
T_{h},& \text{if} ~~~0 \le x \le {L_{0}\over 2};\\
T_{c},& \text{if} ~~~ {L_{0}\over 2} \le x \le L_{0}
\end{array}
\right.
\end{equation}
 as shown in Fig. 1.  The potential  $U_{s}(x)$ and 
$T(x)$ are assumed to be periodic with a period $L_0$, $U_{s}(x+L_{0})=U_{s}(x)$ and $T(x+L_{0})=T(x)$.
\begin{figure}[ht]
\centering
%\subfigure[Bild a.] % caption for subfigure a
{
    %\label{fig:sub:a}
  \includegraphics[width=8cm]{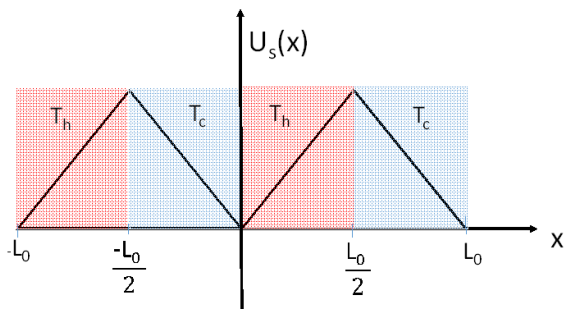}
}
\caption{(Color online) Schematic diagram for a particle that walks  in a piecewise linear potential
in the absence of an external load. }
\label{fig:ratchet}
\end{figure}

For a  fluid such as blood, it is reasonable to
assume that when the temperature of the blood sample  increases by $1.0$ degree Celsius, its viscosity steps down by $2.0$ degree Celsius \cite{aa15}. Thus let us  consider viscous  friction that varies as 
 \begin{equation}
 \gamma(x)=\gamma'-C(T(x)-T_c)
\end{equation}
where $C$  is a constant which is less than  one. In the next two  sections, we will explore the model system in the overdamped and underdamped limits.

\subsection{ Underdamped case in the absence of ratchet potential  }
In this section, we consider   an important model system where a colloidal particle  that  undergoes a biased random walk in  a spatially varying thermal arrangement in the  presence of  external load  $f$ with no potential. 
Solving Eq. (3)  at steady state, the general expression for the probability distribution is obtained as
\begin{equation} 
P(x,v)=\frac{e^{-\frac{m (f-(\gamma+c T_c) v+c v T[x])^2}{2 T[x] (\gamma+c T_c-c
T[x])^2}} \sqrt{\frac{m}{T[x]}}}{\sqrt{2 \pi }}.
   \end{equation}
 The average velocity is found to be  
\begin{equation} 
%v={2fL_{0}+(T_c-T_h)\over 2\gamma L_0}.
v={f\over \gamma(x) }.
\end{equation}
 In the absence of force, the velocity of the particle approaches zero.

Via  Eqs.  (9) and (11),  the entropy production and extraction rates  are calculated as 
 \begin{eqnarray} 
{\dot h}_{d}(t)&=&{\dot e}_{p}(t) \nonumber \\
&=&\frac{1}{2} f^2 L_{0} \left(\frac{1}{\gamma T_c}+\frac{1}{(\gamma+c (T_c-T_h)) T_h}\right)
   \end{eqnarray}
One can see that in the limit where the load approaches the stall  force, ${\dot h}_{d}(t)={\dot e}_{p}(t)=0$.  	Exploiting Eq. (62), it is evident that ${\dot e}_{p}(t)$ and  ${\dot h}_{d}(t)$ increase when the  temperature difference between the hot and cold baths decreases. This is feasible  since when the temperature difference between the heat baths  steps up, the magnitude  of the viscose  friction decreases. 
	
	The rate of heat dissipation 
is calculated  employing  Eq. (20) and it converges to
 \begin{eqnarray} 
{\dot H}_{d}(t)&=&{\dot E}_{p}(t)\frac{1}{2} f^2 L_{0} \left(\frac{1}{\gamma}+\frac{1}{\gamma+c (T_c-T_h)}\right).
   \end{eqnarray}
	In the limit where the load approach  zero,
	${\dot H}_{d}(t)={\dot E}_{p}(t)=0$ showing that at quasistatic limit the system  is reversible.
	On the other hand, the rate of work done is given by 
\begin{eqnarray} 
{\dot W}(t)&=&{\dot E}_{p}(t) \nonumber \\ &=&\frac{1}{2} f^2 L_{0} \left(\frac{1}{\gamma}+\frac{1}{\gamma+c (T_c-T_h)}\right).
   \end{eqnarray}
	For isothermal case $T_{h}=T_{c}$, one gets $v=f/\gamma$, ${\dot h}_{d}(t)={\dot e}_{p}(t)=f^2L_{0}/\gamma T_{c}$ and ${\dot H}_{d}(t)={\dot E}_{p}(t)=f^2L_{0}/\gamma$.

\subsection{ Overdamped case  }

In the presence of ratchet potential, in the overdamped limit,  the closed-form expression for the steady-state current can be given as
\begin{equation}
J= -{\varsigma_{1}\over \varsigma_{2}\varsigma_{3}+(\varsigma_{4}+\varsigma_{5})\varsigma_{1}}
\end{equation}
where the expressions  for $\varsigma_{1}$, $\varsigma_{2}$, $\varsigma_{3}$, and $\varsigma_{4}$
are given by
\begin{eqnarray}
\varsigma_{1}& = &-1+e^{{L_0(f-{2U_{0}\over L_0})\over 2T_{c}}+{L_0(f+{2U_{0}\over L_0})\over 2T_{h}}}\\ \nonumber
\varsigma_{2}& = &{e^{-{{fL_{0}(T_{c}+T_{h})\over T_{c}}+2U_{0}\over 2T_{h}}}\left(e^{fL_{0}\over
                    2T_{c}}-e^{U_{0}\over T_{c}}\right)L_0\over fL_{0}-2U_{0}}\\ \nonumber
          &&-{\left(e^{-{fL_{0}+2U_{0}\over 2T_{h}}}-1\right)L_{0}\over fL_{0}+2U_{0}} \\ \nonumber
\varsigma_{3} & = &{e^{-{U_{0}\over T_{c}}+{fL_{0}+2U_{0}\over 2T_{h}}}
\left(e^{fL_{0}\over 2T_{c}}-e^{U_{0}\over T_{c}}\right)L_0T_{c} \gamma\over fL_{0}-2U_{0}}+\\ \nonumber
            &&{\left(e^{fL_{0}+2U_{0}\over 2T_{h}}-1\right)L_{0}T_{h}(\gamma +c(T_c-T_h))\over fL_{0}+2U_{0}}\\ \nonumber
\varsigma_{4}& = &(\gamma +c(T_c-T_h)){L_{0}^2\left(fL_{0}+2((-1+e^{-{fL_{0}+2U_{0}\over 2T_{h}}})T_{h}+U_{0})\over
                  2(fL_{0}+2U_{0})^2\right)}
\end{eqnarray}
and $\varsigma_{5}=L_{0}^2(t_1+t_2+t_3t_4)$. Here $t_1$, $t_2$, $t_3$ and $t_4$ are given by 
  \begin{eqnarray}
  t_1&=&{\gamma \over 2fL_{0}-4U_{0}}\\ \nonumber
  t_2&=&{\gamma T_c(-1+e^{-{{fL_{0}-2U_0}\over 2T_{c}}})T_c\over (fL_{0}-2U_{0})^2}\\ \nonumber
  t_{3}&=&{(-1+e^{-{{fL_{0}-2U_0}\over 2T_{h}}})T_h\over (f^2L_{0}^2-4U_{0}^2)} \\ \nonumber
  t_{4}&=&{{e^{-{fL_{0}(T_{c}+T_{h})\over 2T_{c}T_{h}}-{U_{0}\over T_{h}}}(e^{{fL_{0}\over
  2T_{c}}}-e^{{U_0\over T_{c}}})(\gamma +c(T_c-T_h))}\over (f^2L_{0}^2-4U_{0}^2)}.
  \end{eqnarray}
	The steady   state  current   converges to zero (at  quasistatic limit) when 
\begin{eqnarray}
f'= {2U_{0}(T_{h}-T_{c})\over (L_{0}(T_{h}+T_{c}))}.
\end{eqnarray}
\begin{figure}[ht]
\centering
%\subfigure[Bild a.] % caption for subfigure a
{
    %\label{fig:sub:a}
    \includegraphics[width=6cm]{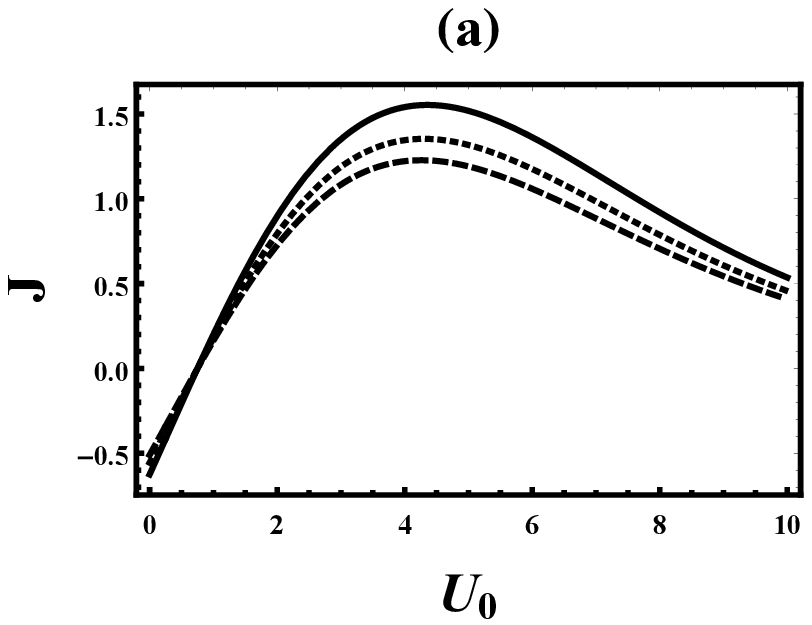}}
\hspace{1cm}
%\subfigure[Bild b.] % caption for subfigure b
{
    %\label{fig:sub:b}
    \includegraphics[width=6cm]{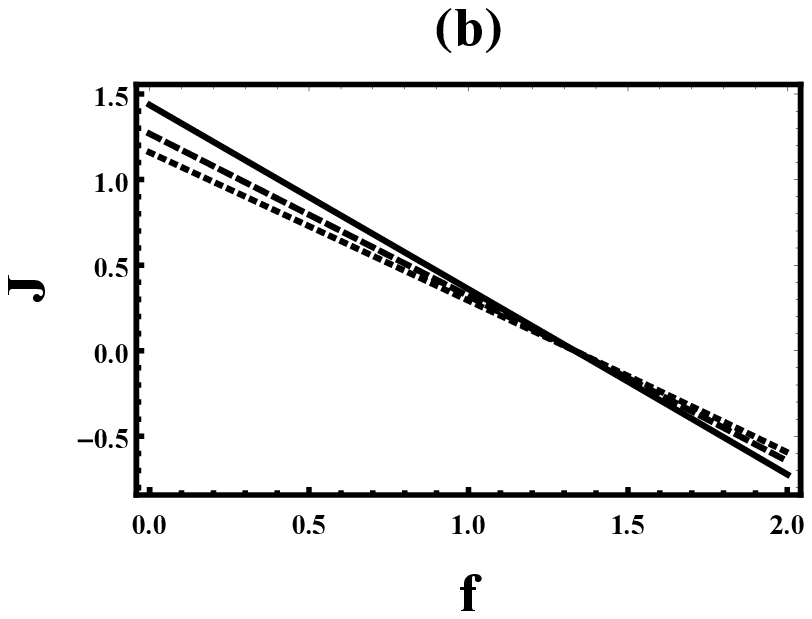}
}
\caption{ (Color online) (a) The dependence of  $J$   on  $U_{0}$ for  fixed     $\tau=2.0$, $\gamma'=1$ and  $f=0.5$. The parameter $C$ is also  fixed as $0.4$ (solid line), $0.2$ (dashed line) and $0.04$ (dotted line). 
 (b)  The plot  $J$      as a function of $f $ for parameter choice $U_{0}=2.0$  and   $\tau=2.0$. 
  The parameter $C$ is  fixed as $0.4$, (solid line), $0.2$ (dashed line) and $0.04$ (dotted line). }
\label{fig:sub} % caption for the whole figure
\end{figure}
Next, let us explore the dependence for the thermodynamic quantities  on the model parameters. In Fig. 4a, the  current as a function of potential height is plotted. The current  exhibits a maximum value at a particular barrier height. As shown in Fig. 4b,   the current  monotonously  decreases with the load.  When $f<f'$, $J>0$  while  when $f>f'$, $J<0$.

Once the expression for steady-state current is obtained, the values for ${\dot h}_{d}$ and   ${\dot e}_{p}$ can  be readily evaluated via Eq. (42). 
At steady state ($v {dv\over dt}=0$), to both underdamped and overdamped cases,  one finds 
\begin{eqnarray}
{\dot h}_{d}  ={\dot e}_{p}
&=&\int \left({J U'(x) \over T(x)} \right)dx.
\end{eqnarray}
The rate of heat extraction  is given by 
\begin{eqnarray}
{\dot H}_{d}=&=&\int \left({J U'(x)} \right)dx.
\end{eqnarray}
At quasistatic limit ( $f\to f'$), ${\dot h}_{d}  ={\dot e}_{p}=0$ as well as ${\dot H}_{d}=0$ since at this limit $J=0$.
\begin{figure}[ht]
\centering
%\subfigure[Bild a.] % caption for subfigure a
{
    %\label{fig:sub:a}
    \includegraphics[width=6cm]{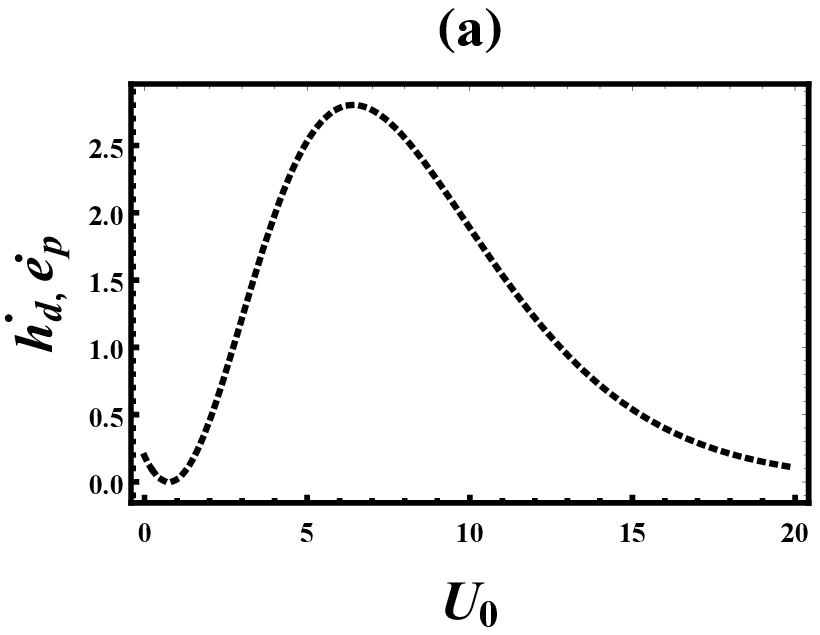}}
\hspace{1cm}
%\subfigure[Bild b.] % caption for subfigure b
{
    %\label{fig:sub:b}
    \includegraphics[width=6cm]{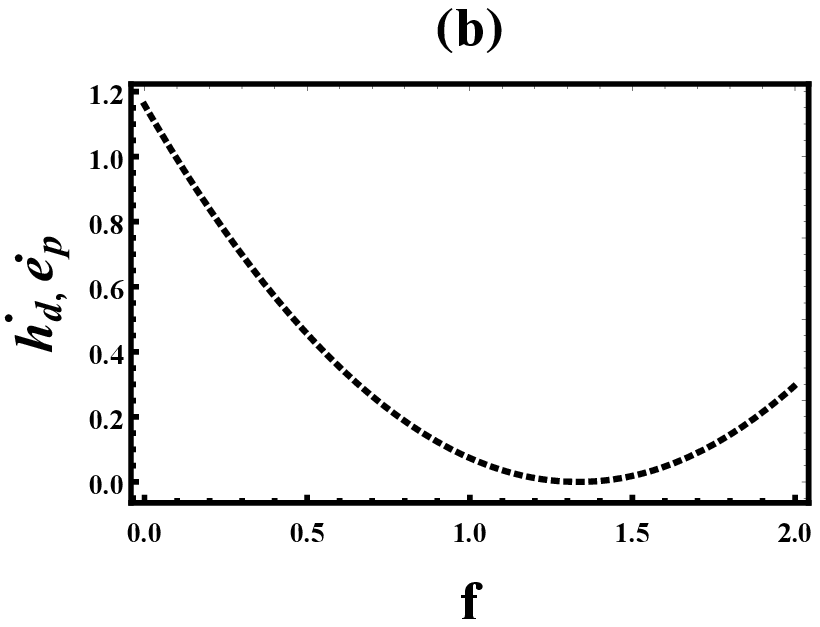}
}
\caption{ (Color online) (a) The plot for   ${\dot e}_{p}(t)$   and ${\dot h}_{d}(t)$  as a function of   $U_{0}$ for parameter choice of $\tau=2.0$, $C=0.04$ and   $f=0.5$. 
 (b)  The plot  ${\dot e}_{p}(t)$    and ${\dot h}_{d}(t)$  as a function of $f$ for parameter choice of  $U_{0}=2.0$, $C=0.04$ and   $\tau=2.0$. } 
\label{fig:sub} % caption for the whole figure
\end{figure}
Let us   explore   how the rate of entropy production ${\dot e}_{p}(t)$ and the rate of entropy extraction   ${\dot h}_{d}(t)$ behave. The plot of  ${\dot e}_{p}(t)$  and ${\dot h}_{d}(t)$  as a function of  $U_{0}$  is depicted in Fig. 5a. The entropy production  and extraction rates take a zero value at the stall force (zero velocity),  ${\dot e}_{p}(t)={\dot h}_{d}(t)=0$  which implies that at the  stall force the system is reversible.  The plot  ${\dot e}_{p}(t)$    and ${\dot h}_{d}(t)$  as a function of $f$ is depicted in  Fig. 5b. As depicted in the figure, the entropy production  and extraction rates  decrease as the load increases and attains a zero value at  the stall force. As the load further increases, ${\dot e}_{p}(t)$ and ${\dot h}_{d}(t)$ step up. 
The  entropy production and  extraction   rates   increase as $C$   and  the temperature  difference between the two baths decreases.  On the other hand,  entropy production  and extraction rates  decrease as $\tau$  increases and attain a zero value at  a particular $\tau$. As the temperature  further increases, ${\dot e}_{p}(t)$ and ${\dot h}_{d}(t)$ increase

If one considers a periodic boundary condition  at steady state in the absence of ratchet potential $U_{0}=0$,  the results obtained quantitively agree with the underdamped case (Section IV  ) and one gets 
	\begin{eqnarray} 
{\dot h}_{d}(t)&=&{\dot e}_{p}(t) \nonumber \\
&=&\frac{1}{2} f^2 L_{0} \left(\frac{1}{\gamma T_c}+\frac{1}{(\gamma+c (T_c-T_h)) T_h}\right)   \end{eqnarray}
	and 
\begin{eqnarray} 
{\dot H}_{d}(t)&=&{\dot E}_{p}(t) \nonumber \\ &=&\frac{1}{2} f^2 L_{0} \left(\frac{1}{\gamma}+\frac{1}{\gamma+c (T_c-T_h)}\right).
   \end{eqnarray}

Let now  explore  the energetics of the model system. When the Brownian particle   along the reaction coordinate, the heat that is taken from the hot heat bath $Q_{h}$  is given as 
\begin{eqnarray}
Q_{h} &=&U_{0}+{fL_{0}\over 2}
  \end{eqnarray}
	while 
	the rate of heat flow into the cold heat bath $Q_{h}$  can be found as 
	\begin{eqnarray}
Q_{c} &=&U_{0}-{fL_{0}\over 2}
  \end{eqnarray}
	which implies the  work done  is  given by 
	\begin{eqnarray}
W &=& Q_{h}-Q_{c}= fL_{0}.
 \end{eqnarray}

Let us now  explore how the efficiency $\eta$ and the coefficient of performance of  the refrigerator $P_{ref}$  behave.  When the engine acts as a heat engine, the efficiency  is given by
\begin{eqnarray}
\eta= {W\over Q_{h}}={fL_{0}\over U_{0}+fL_{0}/2}.
\end{eqnarray}
At quasistatic limit, plugging Eq. (76) into Eq. (84), one gets 
\begin{eqnarray}
\eta= 1-{T_{c}\over T_{h}}
\end{eqnarray}
which is  the efficiency of the Carnot heat engine.  When the  engine performs as a refrigerator, the coefficient of performance of  the refrigerator $P_{ref}$ is given by 
\begin{eqnarray}
P_{ref}= { Q_{c}\over W}={U_{0}-fL_{0}/2 \over fL_{0}}
\end{eqnarray}
and at  quasistatic limit, plugging Eq. (76) into Eq. (86), $P_{ref}$ approaches Carnot refrigerator
\begin{eqnarray}
P_{ref}= {T_{c}\over T_{h}-T_{c}}.
\end{eqnarray}

\section{ Multiplicative noise }
Most of the previous works have focused on calculating the thermodynamic features of different model systems by considering additive noise. Most realistic
 systems  such as neuron  system   can be also described by Langevin
equations with multiplicative noise were in this case,  the noise
amplitude varies spatially \cite{mar12}. Considering multiplicative noise, 
the  intrinsic noise-induced ordering phase transition     has been  also studied in the work  \cite{mar10}.  
In this work, we  study  how entropy, entropy production, and extraction rate depend  on the strength  of the background noise by  solving the model exactly.

For the case where    the temperature is position-dependent $T(x)= \sqrt{D}|x|^{{-z \over 2}}$, in the absence any external potential,   the corresponding Fokker Planck equation is given as 
\begin{eqnarray}
{\partial P(x,t)\over \partial t}&=&{\partial  \over \partial x}\left({T'(x))\over 2 \gamma}\right)P(x,t)+\nonumber \\
&&{\partial  \over \partial x}\left({T(x)\over \gamma}{\partial P(x,t)\over \partial x}\right).
\end{eqnarray}
The probability current is given as 
\begin{eqnarray}
J&=&-\left({P(x,t)T'(x))\over 2 \gamma}\right)-\nonumber \\
&&\left({T(x)\over \gamma}{\partial P(x,t)\over \partial x}\right).
\end{eqnarray}
The solution for the probability distribution is  well known \cite{mulu1} and it is given by  
\begin{eqnarray}
P(x,t) = {|x|^{z \over 2} e^{-{|x|^{z+2}\over D(z+2)^2 t}}\over \sqrt{4\pi D t}}.
\end{eqnarray}
 
From  Eqs. (42) and (43),  one gets 
\begin{eqnarray}
{\dot e}_{p}&=& \int { \gamma J^{2} \over P T(x) } dx 
\end{eqnarray}
and 
\begin{eqnarray}
{\dot h}_{d} &=&\int\left({J(T'(x))\over 2 T(x)}\right) dx. 
\end{eqnarray}

The expression for  the entropy production and extraction rates  can be found by substituting Eqs. (87) and (88) into Eqs.  (89) and (90).
 The plot ${\dot h}_{d}(t)$   and ${\dot e}_{p}(t)$  as a function of $t$ for parameter choice $\tau=1$, $D=1.0$ and  $z=-4.0$ is depicted in Figs. 6a and 6b.  The  figures depicts that ${\dot h}_{d}(t)$ and ${\dot e}_{p}(t)$ decrease  as time increases and in long time limit, it approaches its stationary  value ${\dot h}_{d}(t)={\dot e}_{p}(t)=0$.  Only in the long time limit,  $t\to \infty$, ${d S(t)\over dt}=0$ since  ${\dot e}_{p}(t)= {\dot h}_{d}(t)=0$. This can be intuitively  comprehended  on physical grounds. For  isothermal case, in the long time limit,   the system approaches  stationary state and    only at  this particular  state,  $\Delta h_d=0$,  $\Delta S=0$  or $\Delta e_p=0$ (at stationary state).  However  when the particle operates at finite time, the system operates irreversibility and in this regime,  the second law of thermodynamics  states that $\Delta S(t)>0$. As it can be seen from Fig. 6 that if the thermodynamic quantities are evaluated   in the time interval between $t=0$ and any time $t$, always the inequality 
$\Delta h_d(t)=h_d(t)-h_d(0)>0$,  $\Delta S(t)=S(t)-S(0)>0$  or $\Delta e_p(t)=e_p(t)-e_p(0)>0$ holds true and as time progresses the change in this parameters increases. 
In fact,  in small $t$ regimes,  ${\dot e}_{p}(t)$ becomes much larger than ${\dot h}_{d}(t)$  (see Figs. 6a and 6b) showing  that the entropy production is higher (than entropy extraction)  in  the first few period of times.  When  time increases, more entropy will be  extracted   ${\dot h}_{d}(t)>{\dot e}_{p}(t)$. Over all, since the system produces enormous amount of  entropy at initial time,  in latter time or any time $t$, $\Delta e_p(t)>\Delta h _d(t)$ and hence $\Delta S(t)>0$. 
\begin{figure}[ht]
\centering
%\subfigure[Bild a.] % caption for subfigure a
{
    %\label{fig:sub:a}
    \includegraphics[width=6cm]{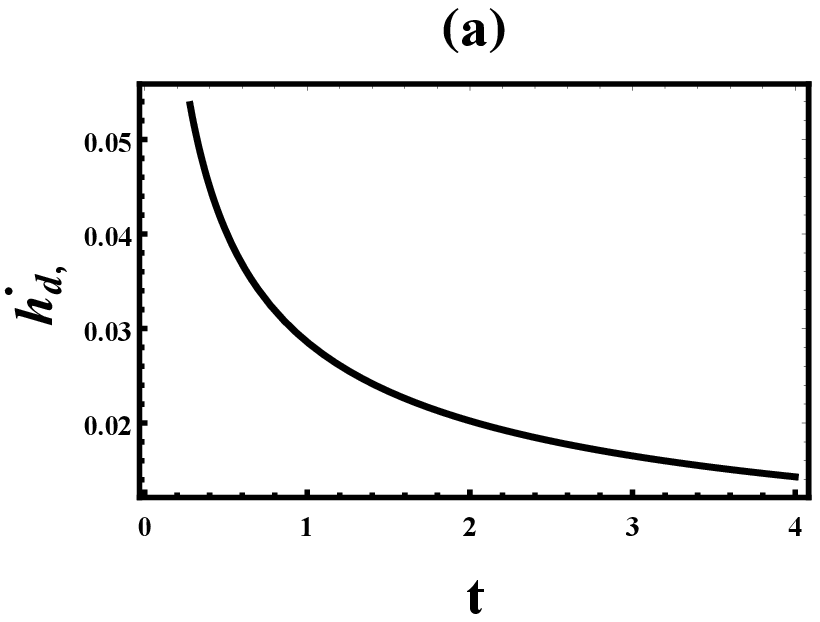}} 
\hspace{1cm}
%\subfigure[Bild b.] % caption for subfigure b
{
    %\label{fig:sub:b}
    \includegraphics[width=6cm]{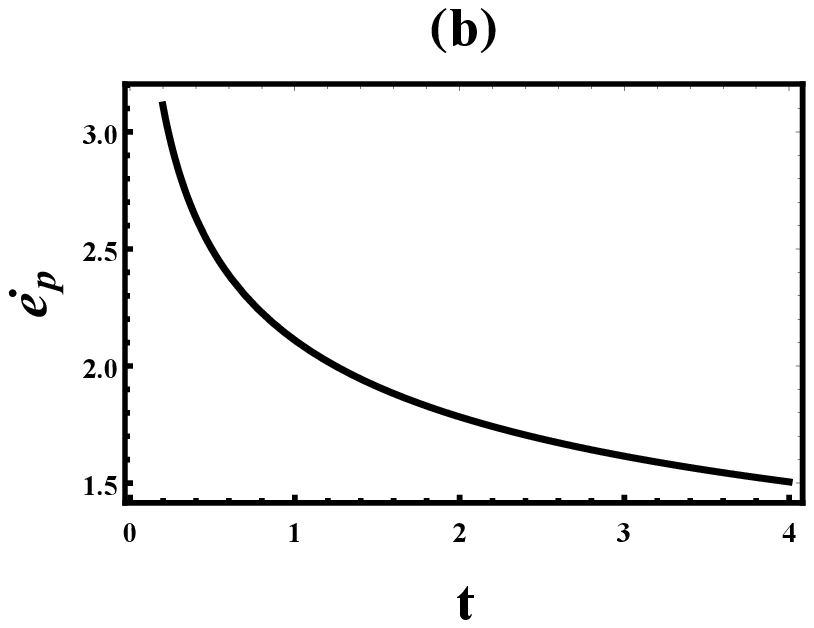}
}
\caption{ (Color online) The plot ${\dot h}_{d}(t)$    and ${\dot e}_{p}(t)$  as a function of $t$ for parameter choice $\tau=1$, $D=1.0$ and  $z=-4.0$ is depicted in Figs. 6a and 6b, respectively.  The  figures depicts that ${\dot h}_{d}(t)$ and ${\dot e}_{p}(t)$ decrease  as time increases and in long time limit, it approaches its stationary  value ${\dot h}_{d}(t)={\dot e}_{p}(t)=0$.  
} 
\label{fig:sub} % caption for the whole figure
\end{figure}

\section{Summary and conclusion}

The influence of viscous friction on the thermodynamic properties of a Brownian particle   that  walks in overdamped and underdamped media is studied. The viscous friction is considered  to  vary either spatially or  temporally. 
By extending  Seifert stochastic approach  to underdamped and overdamped  media,  the general expressions   for  entropy  production,    free energy,  and    entropy extraction rates are derived. To explore 
the non-equilibrium thermodynamic features of   a Brownian particle that  hops  in  medium where  its  viscosity  depends on time,  a  Brownian particle that walks on a periodic isothermal medium (in the presence or absence of load) is considered. The  analytical results depict that  in the absence of  load,  the entropy production rate balances the entropy extraction rate which is  reasonable since any  system which is in contact with a uniform  temperature should obey the detail balance condition in a long time limit.  It is shown that when  a distinct temperature difference is not retained  between the hot and cold  baths, in  absence of load,  the entropy production  still balances  the entropy extraction rate revealing the system is reversible.  When the external load is zero and 
when the viscous friction  decreases  in time,  the entropy  monotonously increases   with time and saturates to  a constant value as $t$ further steps up.  The entropy production rate  decreases in time and at steady state (in the presence of load), ${\dot e}_{p}={\dot h}_{d}>0$ which agrees with the results shown in the works \cite{muuu177}.  On the contrary,  when the  viscous friction  increases in time,  the rate of entropy production as well as the rate of entropy extraction monotonously    steps up   showing that such systems are  inherently irreversible.

For a system where the viscous  friction of a  medium tends to decrease as the temperature of the  medium increases,  the non-equilibrium thermodynamic features of   the model system are explored. In this case, the load $f$  dictates the direction of the particle velocity. The  steady-state velocity of the engine is positive   when $f$ is smaller and the engine acts as a heat engine. In this regime the entropy production and extraction rates become nonzero.  When $f$ steps  up, the velocity of the particle steps down and at stall force, the entropy production rate balances the entropy extraction rate revealing   the system is reversible  at this particular choice of parameter.  For large loads, the current is negative and the engine acts as a refrigerator. In this region the entropy production and extraction rates become nonzero. In the  absence of load, the entropy production and extraction rates become larger  than zero  as long as a  distinct temperature difference is retained between the hot and cold baths. We further  explore  the thermodynamic  features of such systems   by considering a multiplicative noise wherein
case the noise amplitude varies spatially.

In conclusion,   in this work,   we derive  several thermodynamic relations  to a Brownian particle moving in underdamped and overdamped media   by considering viscous friction that varies temporally and spatially. We believe that the  present theoretical work serves as a basic  tool to understand the nonequilibrium  thermodynamics.

\section*{Acknowledgment}
I would like to thank Blaynesh Bezabih and Mulu  Zebene for their
constant encouragement.

\end{document}